\documentclass[aps, prb, reprint, superscriptaddress]{revtex4-1}
\usepackage{amsmath,amssymb,bm}
\usepackage{graphicx}
\newcommand{\Fig}[1]{Fig.\ \ref{fig:#1}}
\begin{document}
\title{Possibility of Deconfined Criticality in SU($N$) Heisenberg Models at Small $N$}
\author{Kenji Harada}
\affiliation{Graduate School of Informatics, Kyoto University, Kyoto 615-8063, Japan}

\author{Takafumi Suzuki}
\affiliation{Graduate School of Engineering, University of Hyogo, Himeji 671-2280, Japan}

\author{Tsuyoshi Okubo}
\affiliation{Institute for Solid State Physics, University of Tokyo, Kashiwa 5-1-5, Chiba 277-8581, Japan}

\author{Haruhiko Matsuo}
\affiliation{Research Organization for Information Science and Technology, Kobe 650-0047, Japan}

\author{Jie Lou}
\affiliation{Department of Physics, Fudan University, Shanghai 200433, China}

\author{Hiroshi Watanabe}
\author{Synge Todo}
\author{Naoki Kawashima}
\affiliation{Institute for Solid State Physics, University of Tokyo, Kashiwa 5-1-5, Kashiwa, Japan 277-8581}
\date{\today}

\begin{abstract}
  To examine the validity of the scenario of the deconfined critical
  phenomena, we carry out a quantum Monte Carlo simulation for the
  SU($N$) generalization of the Heisenberg model with four-body and
  six-body interactions.  The quantum phase transition between the
  SU($N$) N\'eel and valence-bond solid phases is characterized for
  $N=2,3,$ and $4$ on the square and honeycomb lattices.  While
  finite-size scaling analysis works well up to the maximum lattice
  size ($L=256$) and indicates the continuous nature of the phase
  transition, a clear systematic change towards the first-order
  transition is observed in the estimates of the critical exponent $y
  \equiv 1/\nu$ as the system size increases.  We also confirm the
  relevance of a squared valence-bond solid field $\Psi^2$ for the SU(3)
  model.
\end{abstract}
\maketitle

In this Rapid Communication, we consider a quantum phase transition 
that is presumably described by 
the scenario of the {\em deconfined critical phenomena} (DCP)
\cite{SenthilVBSF2004,SenthilBSVF2004,SenthilBSVF2005}.
The DCP transition takes place between the N\'eel state and the 
valence-bond solid (VBS) state in two dimensions.
It is remarkable that the symmetry group of neither phase is 
a subgroup of the other.
This novel property is an outcome of the intrinsic lattice symmetry
and an effect of the Berry phase of which no direct counterpart exists 
in classical critical phenomena.
It was also argued that such a phase transition is realized in some
frustrated magnets\cite{J1J2Paper} as well as superconductors\cite{SuperConductors}.
The Heisenberg model generalized to the SU($N$) symmetry is a candidate 
for a minimal model that realizes such a quantum criticality.
In the square lattice case, while its ground state for $N\le 4$ is the SU($N$) N\'eel state,
for $N\ge 5$ it is the columnar VBS state\cite{ReadS1989a,HaradaTK1998,HaradaKT2003}.
Interestingly, the VBS ground state of the SU($N$) Heisenberg model 
with $N \ge 5$ shows an approximate $U$(1) degeneracy,\cite{KawashimaT2007}
suggesting that the model is close to the critical point described by the DCP scenario.
In the square lattice $J$-$Q$ model that includes an additional four-body spin 
interaction\cite{Sandvik2007}, one can take the model to its exact transition point
by tuning the amplitude of the new term.
The quantum Monte Carlo (QMC) results of the SU(2) $J$-$Q$ model on the square lattice
\cite{Sandvik2007, Melko:2008ip, Sandvik:2010hg}
showed a good finite-size scaling (FSS),
indicating the critical nature of the transition.
Another set of QMC studies\cite{LouSK2007}
not only showed good FSS plots, but also some universal properties
shared by two models with different multibody interaction terms.
Based on these and additional numerical results for $N\ge 5$, 
it was pointed out \cite{KaulS2012} that the $N$ dependence of 
the critical indices is consistent with the $1/N$ expansion.
To examine more directly the effect of the Berry phase, which is
crucial to the DCP scenario,
the bilayer SU($N$) Heisenberg model was studied\cite{Kaul2012}, 
and the results showed that the N\'eel-VBS quantum phase transition for $N \ge 5$ becomes clearly of the
first order upon introduction of the interlayer couplings.
However, a numerical work on the (2+1) dimensional noncompact CP$^1$ model
\cite{KuklovMPST2008}, 
which is believed to belong to the same universality class,
suggested that the transition could be accompanied by a very small 
but finite discontinuity.
The SU(2) $J$-$Q$ model on the square lattice was directly compared\cite{ChenETAL2013}
with the noncompact CP$^1$ model. It showed that there is a range of length scale 
in which the two models have the same behavior, while 
they depart from each other above a certain length scale.
It was argued that the deviation indicates the nonuniversality of the
transition, perhaps a first-order nature, although the possibility
of another fixed point was not completely ruled out.

With all the arguments and evidence, the issue has not been completely
settled.  Below we present results of a systematic study of the SU($N$)
$J$-$Q$ model obtained with a large scale QMC simulation.  Namely, we
have carried out QMC simulations on the SU(2), SU(3), and SU(4) models
not only on the square lattice but also on the honeycomb lattice.  We
have also extended the system size ranges up to $L=256$ and $96$ in
the square and the honeycomb lattices, respectively. Our results do
not put an end to the controversy, but instead they show (i) apparent
universal critical behaviors shared by these lattice models within a
limited size range up to $L\sim 96$, (ii) a clear systematic change towards
the first-order transition in the estimates of
critical indices as a function of the system size in the SU(2) and
SU(3) cases, and (iii) relevance of the $\Psi^2$ scaling field in the
SU(3) case with the assumption of the criticality.

\begin{figure}[tb]
  \begin{center}
  \includegraphics[width=0.43 \textwidth]{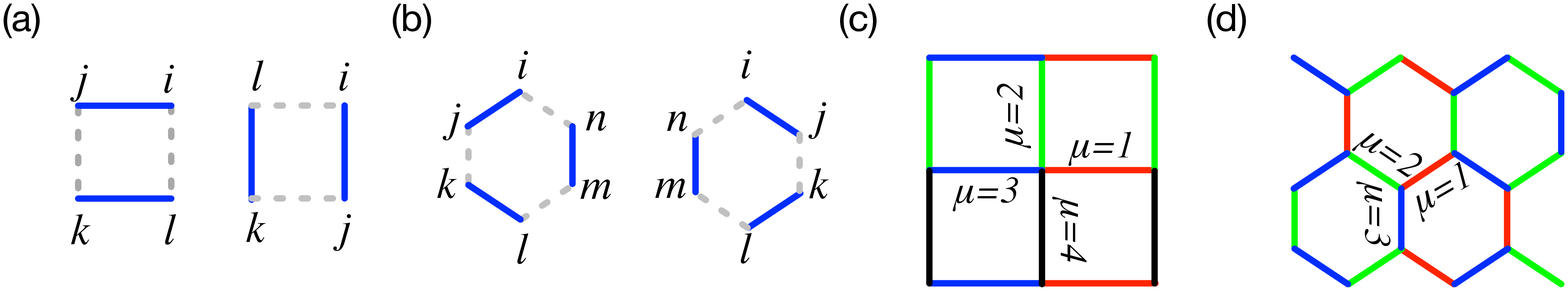}
  \caption{ \label{fig:VBS}   (Color online) The dimer coverings of (a) unit square and (b) unit hexagon in $Q$ terms
  and the classification of the bonds in the definition of 
  the complex order parameter $\Psi_r$, characterizing the VBS phase
  on (c) the square and (d) the honeycomb lattice.}
  \end{center}
\end{figure}

\textit{Model.} The SU($N$) $J$-$Q$ Heisenberg model on the square lattice
can be expressed most simply 
in terms of the projector to the color-singlet state $P_{ij}$:
$$
  H = - J \sum_{(ij)} P_{ij} - Q \sum_{(ij)(kl)} P_{ij} P_{kl},
$$
where the summation in the second term runs over 
all pairs of bonds nearest neighbor and parallel to each other.
Namely, the summation over $(ij)(kl)$ includes two kinds of dimer coverings
of each unit square [\Fig{VBS}(a)].
Likewise, the model on the honeycomb lattice is defined by including
the six-body interaction instead of four, i.e.,
$
  - Q \sum_{(ij)(kl)(mn)} P_{ij} P_{kl} P_{mn}
$
for all unit hexagons, and the summation over $(ij)(kl)(mn)$ includes two kinds of dimer coverings for each hexagon [\Fig{VBS}(b)].
Thus the $Q$ term does not break the lattice rotational symmetry in either case.
The color-singlet projector is expressed 
in terms of the generators of the SU($N$) algebra as
$
  P_{ij} = - \frac1N \sum_{\alpha=1}^N\sum_{\beta=1}^N
             S_i^{\alpha\beta} S_j^{\beta\alpha}.
$
We adapt the fundamental representation on one sublattice
and the conjugate representation on the other sublattice.
Our QMC simulation is based on 
the world-line representation with the loop update
\cite{EvertzLM1993} and its generalization to the SU($N$) model
\cite{HaradaKT2003}.
We used the modified ALPS/LOOPER code\cite{todoK2001, ALPS} with parallelization\cite{TodoMS2012}.
In approaching the quantum critical point, 
we set the inverse temperature $\beta$ as $\beta J = L$,
with the periodic square (rhombic) boundary condition 
on the square (honeycomb) lattice.
In the FSS analysis,
we used the Bayesian method\cite{Harada:2011js} for obtaining
the parameters.

\begin{figure}[tb]
  \begin{center}
  \includegraphics[width=0.45 \textwidth]{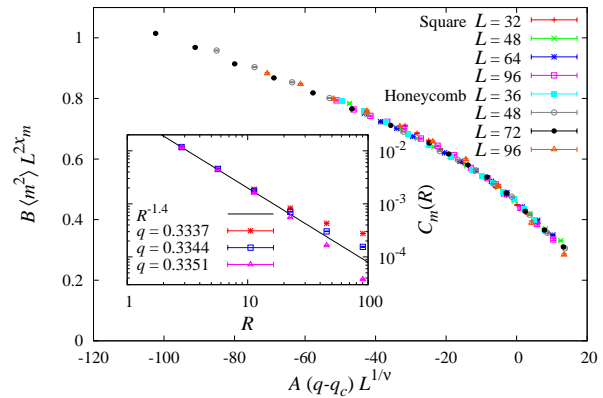}
  \caption{ \label{fig:MFSS}
  (Color online) The FSS plot of the magnetization for the SU(3)
  $J$-$Q$ model on the square and the honeycomb lattices with the system
  size being restricted to $L\le 96$.  The scaling factors $A=2.218$
  and $B=1.752$ are multiplied to the horizontal axis and the vertical
  axis, respectively for the results of the honeycomb lattice ($A=B=1$ for
  the square lattice.) Inset: The two-point correlation function of
  magnetic order for the SU(3) models on a square lattice of $L=256$
  as a function of the distance at various $q$ near the critical
  point.  The straight line corresponds to the estimate of $2x_m$ from
  the squared magnetization. }
  \end{center}
\end{figure}

\textit{Scaling of the magnetic order parameter.} We define the ``magnetic order parameter'' as
$
  m \equiv \frac1{V}\sum_i (S_i^{\alpha\alpha}-1/N) = \frac1{V} \sum_i m_{i,\alpha},
$
with $V$ being the total number of sites,
which agrees with the staggered magnetization in the SU(2) case.
We try to fit the obtained estimates of the magnetization 
to the FSS form
$
  \langle m^2 \rangle = L^{-2x_m} \tilde \chi( t L^y )
$
with $t \equiv q - q_c$ and $y \equiv 1/\nu$, where $q \equiv Q / (J + Q)$.
Figure \ref{fig:MFSS} shows the result of the fitting with limited system sizes
$L\le 96$ in the case of the SU(3) model on both the square and honeycomb lattices.
(This system-size range is simply due to the fact that
the largest system on the honeycomb lattice studied here is $L=96$.)
The values used for \Fig{MFSS} are $q_c = 0.3353$ for the square lattice
and $q_c = 0.2036$ for the honeycomb lattice.
The same values of exponents 
$y=1.87$ and $2x_m=1 + \eta_m = 1.40$ are used for both lattices.
The horizontal and vertical axes for the honeycomb lattice are 
rescaled to match those of the square lattice.
Note that not only do different system sizes fall on the same curve,
but also two lattices with different lattice rotational symmetries also 
collapse on the same curve.
Since the symmetry that is broken at the transition point is different
in the two cases, a different universality class may be naturally expected.
Therefore, this universal behavior supports the DCP scenario.
From a similar analysis we obtain reasonably 
good FSS plots for SU(2) and SU(4) models as well.
The data for the SU(2) model produce
the best fitting with $y = 1.78$ and $2x_m = 1.289$
at $q_c = 0.9585$ and $q_c = 0.5440$ 
for the square and the honeycomb lattice, respectively.
Note here that we use data for $L \le 128$ and $96$ for the square and honeycomb lattices, respectively.
These values are consistent with a recent parallel work that reports 
the DCP on the SU(2) honeycomb lattice model\cite{Pujari:2013df}.
For the data for the SU(4) model with $L \le 96$,
the best fitting is obtained with $y = 1.59$ and $2x_m = 1.486$
at $q_c = 0.0829$ and $q_c = 0.0154$ 
for the square and the honeycomb lattice, respectively.
The inset of \Fig{MFSS} shows the two-point correlators 
$C_m(R_{ij}) \equiv \frac1{N} \sum_{\alpha} 
\langle m_{i,\alpha} m_{j,\alpha} \rangle$
as a function of the distance for the SU(3) models on the square lattice of $L=256$.
The critical value estimated from the data of $L \le 256$ is $q=0.3343(1)$.
While it is slightly smaller than the one quoted above for smaller systems, 
the correlation decay up to $R \sim 24$ lattice units is not so sensitive to $q$ 
and is well characterized by the exponent obtained above from the squared magnetization.

\begin{figure}[tb]
  \begin{center}
  \includegraphics[width=0.45 \textwidth]{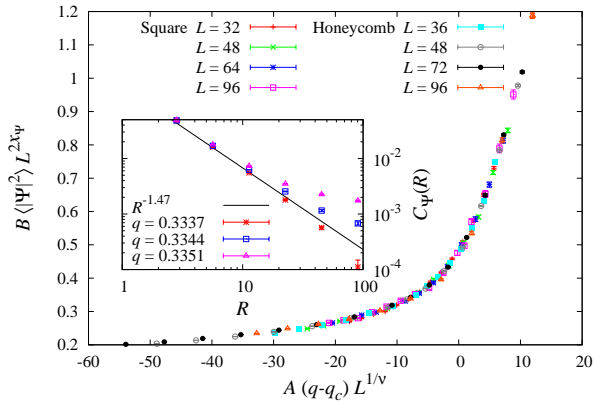}
  \caption{ \label{fig:PSIFSS}
  (Color online) The FSS plot for the squared amplitude of total VBS order
  for the SU(3) $J$-$Q$ model on the square and the honeycomb lattices
  with the system size being restricted to $L\le 96$.
  We set the scaling factors $A=2.422$ and $B=0.6388$ for the honeycomb lattice.
  Inset: The two-point correlation function of
  VBS order for the SU(3) models on a square lattice of
  $L=256$ as a function of the distance at various $q$ near the
  critical point.  The straight line corresponds to the estimate for
  $2x_{\Psi}$ obtained from the squared amplitude of the total VBS order.}
  \end{center}
\end{figure}

\textit{Scaling of the VBS order parameter.} We define the local complex order parameter characterizing the VBS phase as $
\Psi_r \equiv \sum_{\mu=1}^z e^{\frac{2\pi i}{z}\mu}\hat P_{r,r_{\mu}}$
where $z$ is the coordination number of a lattice,
$r_{\mu}$ is the neighboring site of $r$ in the direction $\mu$
[see Figs.~1(c) and 1(d)), and $\hat P_{r,r_\mu}$ is the diagonal part of the projection operator, respectively.
According to Read and Sachdev\cite{ReadS1989a,ReadS1990}, 
the complex VBS operator $\Psi$ corresponds 
to the annihilation operators of skyrmions,
and it is then interpreted \cite{SenthilBSVF2004} in the continuum limit
as $\Psi \sim \psi_{\uparrow}\psi_{\downarrow}^{\ast}$ where $\psi_{\alpha}$ is 
the meron (half-skyrmion) annihilation operator with an up ($\alpha=\uparrow$)
or down ($\alpha=\downarrow$) spin at the core.
In \Fig{PSIFSS}, the FSS of the squared amplitude of the total VBS order
$\langle |\Psi|^2 \rangle \equiv V^{-2}\langle |\sum_r \Psi_r|^2 \rangle$
is presented for the SU(3) model. 
Similar to \Fig{MFSS}, we again restrict the system size to $L\le 96$ and
assume the same critical dimensions for both lattices.
Namely, we assume $y = 1.72$ and $2x_{\Psi} = 1 + \eta_{\Psi} = 1.47$
for both lattices while $q_c = 0.3339$ and $q_c = 0.2029$ for the square 
and honeycomb lattices, respectively.
Note that the value of the scaling exponent $y$ is close 
(and actually consistent within the statistical error)
to the estimate of $y$ based on the magnetization discussed above.
From a similar analysis we obtain a reasonably 
good FSS plot for the SU(2) model
when we assume $y = 1.73$ and $2x_{\Psi} = 1.36$ for the scaling exponents
and $q_c = 0.9552$ and $q_c = 0.5420$ for the square and honeycomb lattices, respectively.
Note here that we again use data for $L \le 128$ and $96$ for the square and honeycomb lattices, respectively.
As for the SU(4) model for $L \le 96$,
the best fitting is obtained with $y = 1.40$ and $2x_{\Psi} = 1.73$ at
$q_c = 0.0805$ and $q_c = 0.0150$ for the square and the honeycomb lattice, respectively.
The two-point correlator
of the local VBS order parameter 
$
  C_{\Psi}(R_{ij}) \equiv
  \langle \Psi^{\ast}_i \Psi_j \rangle
$
is also shown in the inset of \Fig{PSIFSS}.
Again, we observe the consistency between the estimates of the
critical indices obtained from $\langle |\Psi|^2 \rangle$
and $C_{\Psi}$.

\begin{figure}[tb]
  \begin{center}
  \includegraphics[width=0.44 \textwidth]{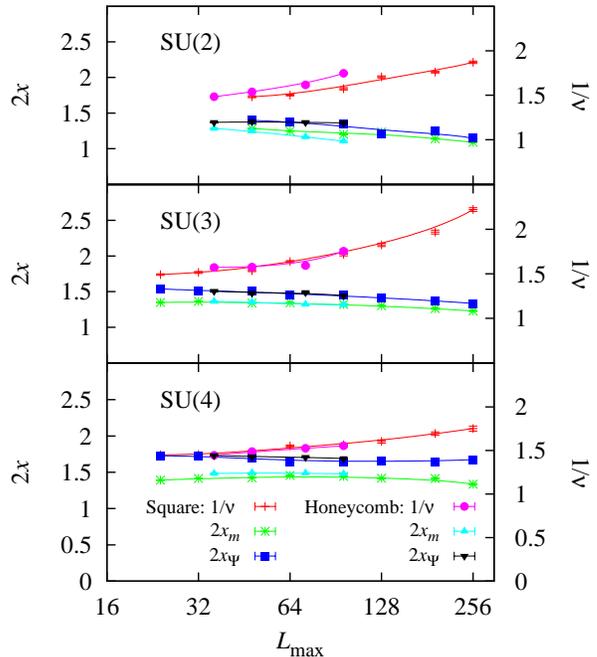}
  \caption{ \label{fig:SizeDependence} (Color online) The estimate of
    the scaling dimensions $y\equiv 1/\nu$ and $2x$ for the SU($N$)
    model on the square and honeycomb lattices as a function of the
    system size used in the FSS. At $L_{\max}$ = 256, $q_c$ = 0.9568(2), 0.3343(1), and 0.0814(3) for SU($N$=2,3,4) models on a square lattice, respectively.}
  \end{center}
\end{figure}

\textit{Systematic size dependence.} The strongest skepticism concerning the critical nature of the phase transition
comes from the argument\cite{KuklovMPST2008} 
that the true nature of the transition is revealed 
only in a very long range behavior, and that the previously attainable
system size might not have reached that regime.
In order to see the systematic trend as we go to a larger length scale,
we apply the FSS analysis for
quadruplets of the system sizes
from $L_{\max}/3$ to $L_{\max}$
such as \{32, 48, 64, 96$(=L_{\max})$\},
and systematically change the value of $L_{\max}$.
In \Fig{SizeDependence},
we plot the $L_{\max}$ dependence of these estimated scaling dimensions.
We use the same value of $q_c$ and $y \equiv 1/\nu$ for both the magnetization and the VBS order parameter.
(This time we have lifted the restriction that $y$ should 
be independent of the lattice.)
As is evident, for the SU(2) and the SU(3) models,
there is a systematic trend of increasing $y$ 
as a function of $L_{\max}$.
Whether it will eventually reach the value $y=d+1=3$, 
the value expected for the first-order transition, 
cannot be judged from the present data.
We should note here that large values of $y$ do not necessarily
suggest the first-order transition since 
the $1/N$ expansion\cite{KaulS2008} also predicts large values
for the DCP fixed point for small $N$.
The systematic decrease in
the effective values of the scaling dimensions 
$2x_m$ and $2x_{\Psi}$ is not as strong as $y$, 
while both of them should converge to zero if the transition is of the first order.
The size dependence of the SU(4) model looks somehow different 
from those of the other cases.
The systematic drifts in the critical indices are much weaker.

\begin{figure}[tb]
  \begin{center}
  \includegraphics[width=0.44 \textwidth]{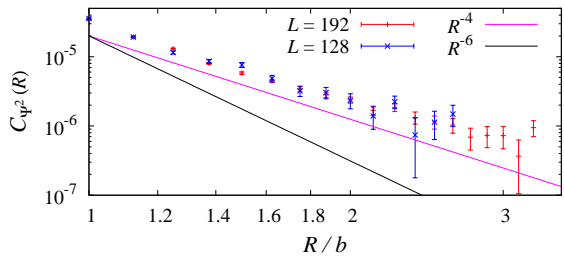}
  \caption{ \label{fig:PSI2COR}(Color online) The two-point correlation function of the square of block-averaged 
order parameter $(\bar{\Psi})^2$
for the SU(3) models as a function of the distance at $q = 0.3343$ which is the critical point
estimated from the data of $L \le 256$.
The block size $b$ is 8. }
  \end{center}
\end{figure}

\textit{Higher power of local VBS order.} We can directly estimate
the order of the quantum VBS phase transition through the
relevance or irrelevance of the higher power of the local VBS order
parameter $\Psi^q$ with $q \ge 2$. In order for the quantum
transition on the square lattice to be of second order, $\Psi^4$
must be irrelevant since this field is naturally included in the
Hamiltonian for the lattice with $Z_4$ symmetry.  Similarly, the
relevance or irrelevance of $\Psi^3$ and $\Psi^2$ is crucial,
respectively, to the transition in the honeycomb lattice and to that
in the square lattice with strong spatial
anisotropy\cite{HaradaKT2007,Grover:2007gf}.  The straightforward
definition of $\Psi^q$ is not suitable here because of the discrete
nature of the present $\Psi$ in contrast to its counterpart in the
continuous field theory.  Instead, we define the block-averaged order
parameter $ \bar\Psi_i \equiv \frac1{b^d} \sum_{j \in \Omega_b(i)}
\Psi_{j}$, where $\Omega_b(i)$ is the region of a square with the
linear scale $b$ centered at $i$.  In \Fig{PSI2COR}, the two-point
correlator for $q=2$, $ C_{\Psi^2}(R_{ij}) \equiv \langle (\bar
\Psi_i^{\ast})^2 (\bar \Psi_j)^2 \rangle $ is plotted against the
distance.  From the figure, we estimate the scaling dimension of
squared VBS order as $2x_{\Psi^2} \sim 4.0$, or $y_{\Psi^2} \sim 1.0 >
0$, i.e., the skyrmion-pair operator is relevant at the deconfined
critical point.  This result shows that even if the DCP scenario is
the correct description of the N\'eel-VBS quantum phase transition, it
cannot take place in the SU(3) $J$-$Q$ model with only $Z_2$ lattice
rotational symmetry, where the model intrinsically contains the
relevant perturbation $\Psi^2$. It follows, for example, that the
quasi-one-dimensional biquadratic Heisenberg model should not have a DCP critical point\cite{HaradaKT2007}.
In previous work\cite{HaradaKT2007}, the numerical results were
analyzed mainly with the assumption of the criticality.  This was
partially due to the very slow onset of the order parameter as a
function of the system size and the absence of a clear discontinuity,
particularly for small systems.  However, a trend similar to the one
described above was already detected there; the large estimates of $y
\equiv 1/\nu$ [e.g., $y = 2.5(2)$ for the VBS order parameter, and $y
= 2.9(2)$ for the magnetization order parameter].  The discontinuous
nature of the spatially anisotropic SU(3) model is also consistent
with the conclusion of a recent study based on the FSS analysis for the same
model \cite{Block:2013bx}.  In the cases of $q \ge 3$, the scaling
dimension is probably larger than that of $q=2$. However, we have
not succeeded in reliably estimating them due to relatively large
statistical noises.

\textit{Conclusions.} We have presented a series of numerical results 
on the SU($N$) $J$-$Q$ models on two lattices, square and honeycomb ones.
Up to the system size explored in the present study, all the scaling analyses work fine
as long as the range of the system size is not broad.
Based on the assumption of the criticality, we have estimated
various critical indices.
The estimates obtained with different quantities on different lattices
are consistent with each other.
In addition, the agreement between the numerical estimates of 
the critical exponents $x_{\Psi}$ and $x_m$ with the 
$1/N$ expansion\cite{KaulS2008} is still good even after the
present updates of the former based on the largest systems. 
These pieces of evidence are consistent with the DCP scenario.
However, the previous estimate of $y$ in Ref.~\onlinecite{LouSK2007} had to be
considerably shifted beyond the error estimated then.
This trend is systematic; for the SU(2) and SU(3) cases at least, 
the estimate increases as the system becomes larger
and it seems to continue to grow beyond the largest estimate obtained
in the present work.
Because of these observations, 
we have to keep the possibility of a first-order transition still open.
If the transition is of first order, the question 
will be about the reason for the apparent universal behaviors.
It would be rather difficult to explain the observed behaviors
unless the DCP fixed point exists close to the renormalization group trajectory
of the present model,
even if it may not be the governing fixed point.
Whether the difference between SU(4) and the other two cases 
persists for larger systems is also an important question 
and requires further studies.

\textit{Acknowledgments.} The computation in the present work was executed on computers at the
Supercomputer Center, ISSP, University of Tokyo, and also on the K
computer at the RIKEN AICS (Project No. hp120283).  The present
work is financially supported by MEXT KAKENHI No. 25287097, and by CMSI, MEXT-SPIRE, Japan.

\bibliography{SUNJQ2013.bib}

\end{document}